\documentclass[conference]{IEEEtran}
\IEEEoverridecommandlockouts
\usepackage{amsmath,amssymb,amsfonts}
\usepackage{amsthm}

\usepackage[ruled,noline,noend,linesnumbered]{algorithm2e}
\SetAlgoNoEnd              %
\DontPrintSemicolon        %
\SetKwFunction{PlaceJobs}{PlaceJobs}
\SetKwFunction{SelectVictims}{SelectVictims}
\SetKwInOut{KwIn}{Input}
\SetKwInOut{KwOut}{Output}
\SetKwComment{Comment}{$\triangleright$\ }{}
\SetCommentSty{textnormal}

\usepackage{array}
\usepackage{booktabs}
\usepackage{multirow}
\usepackage{makecell}

\usepackage{graphicx}
\usepackage{rotating}
\usepackage{stfloats}

\usepackage{textcomp}
\usepackage{xcolor}
\usepackage{xspace}
\usepackage{enumitem}
\usepackage{url}
\usepackage{verbatim}

\usepackage[nocompress]{cite}

\usepackage[
    colorlinks=true,
    citecolor={blue!70!black},
    linkcolor={red!60!black},
    urlcolor={blue!70!black},
    breaklinks=true
]{hyperref}

\def\BibTeX{{\rm B\kern-.05em{\sc i\kern-.025em b}\kern-.08em
    T\kern-.1667em\lower.7ex\hbox{E}\kern-.125emX}}

\newcommand{\SystemName}{DeepShare\xspace}

\newcommand{\mypara}[1]{\smallskip\noindent\textbf{#1}\quad}

\usepackage{eso-pic}

\begin{document}

\title{DeepShare: Assurance-Driven Deep Learning Job Scheduling for Multi-Tenant Clusters}

\author{
    \IEEEauthorblockN{Jinghao Wang\textsuperscript{1}, 
    Yihang Zhou\textsuperscript{1},
    Xiao Zhou\textsuperscript{1},
    Xinlei Zheng\textsuperscript{1}, 
    Xiaoyang Sun\textsuperscript{2\dag},
    Tianyu Wo\textsuperscript{1}, \\
    Chunming Hu\textsuperscript{1},
    Renyu Yang\textsuperscript{1}
    \thanks{\dag  Dr. Xiaoyang Sun is the corresponding author.}}
    \IEEEauthorblockA{
    \textsuperscript{1}Beihang University \quad\quad  \textsuperscript{2}University of Leeds \\
    \{wang\_jinghao, zhou\_yihang, zhouxiao2021, xinleizh, woty, hucm, renyuyang\}@buaa.edu.cn; x.sun4@leeds.ac.uk
    }
}

\maketitle

\AddToShipoutPictureFG*{%
  \AtPageLowerLeft{%
    \put(36,18){\parbox[b]{540pt}{\fontsize{7}{8}\selectfont
      \textcopyright\ 2026 IEEE. Personal use of this material is permitted. Permission from IEEE must be obtained for all other uses, in any current or future media, including reprinting/republishing this material for advertising or promotional purposes, creating new collective works, for resale or redistribution to servers or lists, or reuse of any copyrighted component of this work in other works.
    }}%
  }%
}

\begin{abstract}

Multi-tenant GPU clusters frequently remain underutilized even when tenants experience long queueing delays, because quota control, queue ordering, preemption, and GPU sharing are driven by different local signals. We present \SystemName, a scheduler that uses a continuous tenant-assurance signal to coordinate these decisions at runtime. \SystemName combines elastic quota borrowing, tenant-specific runtime prediction, cost-aware best-effort preemption, and interference-aware MPS colocation, while using the same assurance signal to decide when borrowed capacity should be reclaimed and when sharing should become more conservative. In trace-driven experiments on 23{,}859 Venus jobs and 3{,}200 internal jobs, \SystemName achieves an average GPU utilization of 70.58\%, a 29.5\% improvement over the strongest non-intrusive sharing baseline, while reducing average queueing delay by 46\%. On a 16-GPU Kubernetes testbed, it reduces the average job completion time by 34\% and maintains 93\% QoS compliance for guaranteed tenants. These results show that treating tenant assurance as a runtime control loop achieves a more advantageous utilization-QoS trade-off than optimizing quotas, scheduling, and resource sharing independently.
\end{abstract}

\section{Introduction}
\label{sec:introduction}

Deep learning (DL) training workloads~\cite{yang2025qwen3, liu2024deepseek, touvron2023llama} have become primary consumers of GPU resources in modern data centers, powering applications in computer vision~\cite{he2016deep}, mathematical reasoning~\cite{shao2024deepseekmath}, and scientific research~\cite{merchant2023scaling}. To improve hardware efficiency, industry and academia widely adopt multi-tenant GPU clusters~\cite{verma2015large, xiao2020antman}, where resources are centrally managed across users and teams. In practice, such clusters often host long-running training, short exploratory jobs, and opportunistic experiments, making static resource allocation difficult to keep efficient. In such clusters, unused capacity cannot simply be reassigned without affecting tenant guarantees, yet production deployments still leave large GPU fractions idle or lightly used~\cite{jeon2019analysis, weng2022mlaas}. On Alibaba's PAI platform, more than 75\% of tasks request less than 10\% of a GPU, while average utilization remains only 25--50\%~\cite{weng2022mlaas, weng2023beware}; similar inefficiency appears in large-scale clusters~\cite{sun2018rose}, where compute and memory utilization remain low despite high demand.

This underutilization stems not only from static allocation but also from resource-management decisions made with little awareness of one another. Quota mechanisms assign tenant entitlements, queueing policies choose the next job, and sharing policies decide whether two jobs can safely share a GPU. The question is not only how to find idle GPUs, but also when their use remains safe for future quota recovery. These choices interact: borrowed quota must be reclaimable, short-job priority must not delay under-served tenants, and low-interference colocation may still be undesirable if it slows quota recovery~\cite{jeon2019analysis, weng2022mlaas, zhao2020hived}.

Recent work has improved individual parts of GPU cluster management, including job colocation~\cite{hu2024characterization, strati2024orion}, performance-aware scheduling~\cite{gu2021liquid, narayanan2020heterogeneity, jayaram2023sia}, and fairness-oriented allocation~\cite{zheng2023shockwave}. Yet framework-level sharing mechanisms~\cite{xiao2018gandiva, xiao2020antman} often sacrifice portability, fragmentation-aware strategies rely on static boundaries, runtime prediction approaches~\cite{peng2018optimus, gu2019tiresias} are sensitive to workload regularity, and fairness models provide limited support for per-tenant service differentiation. A locally attractive decision, such as admitting a short job or a low-interference pair, can still be harmful if it delays quota recovery. The missing link is runtime coordination: a scheduler must know not only whether a job is short or a pair has low interference, but also whether admitting it helps or hurts recovery of tenants whose guarantees are unmet.

We introduce \SystemName, a Kubernetes-native resource management framework built around the \emph{Quota Assurance Degree} $\tilde{Q}_i(t)$. QAD measures how far tenant~$i$ is from receiving its guaranteed share over time: high QAD means \textit{guaranteed} demand is served, while low QAD exposes under-service and raises recovery priority. By comparing allocation with the smaller of quota and current \textit{guaranteed} demand, QAD avoids rewarding artificial demand inflation or penalizing temporarily idle quota. Smoothing avoids reacting to short-lived fluctuations while preserving persistent deficits for the scheduler. \textit{Best-effort}\footnote{In this paper, \textit{best-effort} jobs consume reclaimable surplus capacity beyond a tenant's quota and may be preempted when \textit{guaranteed} demand rises.} resources are reclaimed when they interfere with recovery.

\SystemName uses QAD across scheduling decisions. Deficit-aware Reclaimable Allocation (DRA, $\S$\ref{subsec:dra}) exposes idle capacity to \textit{best-effort} workloads, while QAD decides when borrowed capacity gives way to \textit{guaranteed} demand. Runtime prediction ($\S$\ref{subsec:predictive}) is applied after tenant recovery priority, so short jobs are favored only among tenants with comparable assurance states. Interference-aware colocation ($\S$\ref{subsec:colocation}) considers predicted slowdown and tenant assurance, becoming more conservative when either tenant is under-served, without modifying user code or DL frameworks. As a tenant recovers, the same signal relaxes these decisions and re-enables more aggressive sharing.

Evaluation shows that \SystemName increases average GPU utilization to 70.58\% in trace-driven simulations, a 29.5\% improvement over Lucid~\cite{hu2023lucid}, while reducing average queueing delay by 46\%. On a Kubernetes testbed, \SystemName reduces average job completion time by 34\%, queueing delay by 66\%, and maintains 93\% QoS compliance for \emph{guaranteed} tenants~\cite{k8sschedulerplugins, volcano2024}. Ablation shows that elastic borrowing, runtime-aware ordering, and QAD-aware colocation reduce queueing delay by 31\% beyond DRA ($\S$\ref{subsec:dra}) alone.

The main contributions of this paper are:

\begin{itemize}[nosep]
    \item QAD, a continuous tenant-assurance measure that compares recovery urgency across tenants and distinguishes transient quota fluctuations from persistent under-service.

    \item An assurance-aware scheduling policy that applies QAD before runtime prediction, reducing queueing delay without letting short-job optimization override tenant recovery.

    \item A Kubernetes-native scheduler with reclaimable \textit{best-effort} borrowing, cost-sensitive preemption, and QAD-aware MPS colocation, evaluated through trace-driven simulation and a 16-GPU deployment.
    
\end{itemize}

\section{Background and Motivation}
\label{sec:background}

\subsection{GPU Cluster Scheduling and Sharing}\label{subsec:bg-overview}

Multi-tenant GPU clusters have become the primary platform for deep learning (DL) research and deployment~\cite{hu2021characterization, weng2022mlaas, chen2019deep, jeon2019analysis}. For isolation and fairness, they are commonly partitioned into \emph{virtual clusters} (VCs) with fixed resource quotas~\cite{hu2024characterization, ye2024deep}. Fixed quotas make resource allocation predictable, but they also make it difficult to reuse temporarily idle capacity across tenants.

At the same time, advances in GPU compute capability have increased the gap between hardware capacity and the resource demands of many individual DL jobs. GPU sharing~\cite{xiao2018gandiva} addresses this gap at multiple levels---hardware partitioning (e.g., NVIDIA MIG~\cite{nvidia_mig_doc}), compute-level multiplexing (e.g., MPS~\cite{nvidia_mps_doc}), and driver-level software virtualization---each trading off isolation against flexibility. A growing body of schedulers builds on these mechanisms to colocate multiple jobs per GPU~\cite{xiao2018gandiva, xiao2020antman, hu2023lucid, strati2024orion, liu2025smore, liu2025isacpp}, employing time multiplexing, spatial partitioning, or interference-aware placement combined with higher-level scheduling policies~\cite{xiao2018gandiva, hu2023lucid}.

\subsection{Empirical Study of a Multi-Tenant GPU Cluster}\label{subsec:bg-empirical}

\begin{figure}[t!]
    \centering
    \includegraphics[width=\columnwidth]{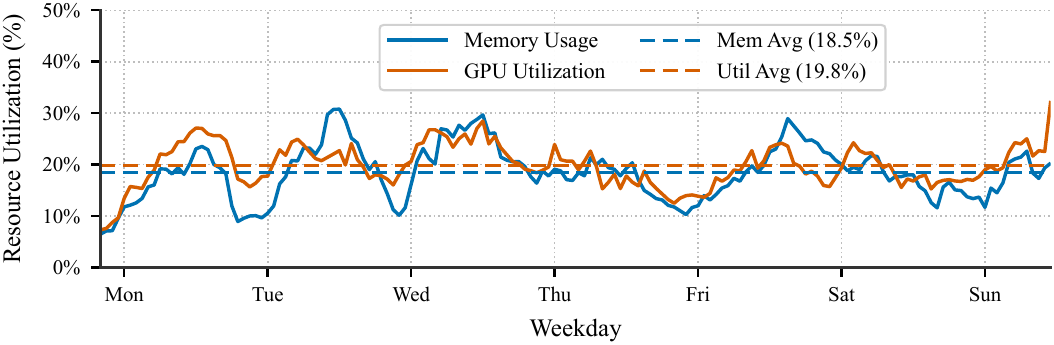}
    \caption{GPU and memory utilization across one week.}
    \label{fig:cluster_utilization}
\end{figure}

Although administratively convenient, fixed quotas introduce a temporal mismatch: they are provisioned on weekly or monthly timescales, while DL workloads are bursty over hours or minutes. This misalignment leads to jobs queueing behind quota limits even as GPUs remain underutilized, as shown in Fig.~\ref{fig:cluster_utilization}. Production studies report overall GPU utilization of only 25\%–52\%~\cite{jeon2019analysis, weng2022mlaas, weng2023beware}.

To inform our design, we profiled a university-managed GPU cluster with 58 nodes and 219 GPUs, including NVIDIA Tesla V100-32GB and A100-40GB devices, partitioned into 12 virtual clusters with fixed monthly quotas. We collected one week of telemetry via NVIDIA DCGM, including GPU-level metrics (e.g., SM utilization, memory bandwidth), job-level statistics (e.g., resource requests), and user-level submission patterns. 

\mypara{Observation 1: Persistent Underutilization.}
Figure~\ref{fig:cluster_utilization} highlights a systematic mismatch between queueing pressure and actual GPU usage. During the seven-day period, average GPU compute utilization is 19.8\% and average memory utilization is 18.5\%. A clear diurnal rhythm emerges: arrivals are concentrated during daytime hours (e.g., 10:00--12:00, 14:00--18:00), with substantially lower submission rates at night. Burst submissions fully consume group quotas, after which allocations remain idle until the next active window. 

\mypara{Observation 2: Heterogeneity and Colocation Potential.}
The workload shows strong heterogeneity along two dimensions. In terms of duration, short exploratory jobs (median runtime under 30 minutes) constitute 62\% of submissions but use only 11\% of GPU-hours, whereas long training jobs (over 4 hours) consume 73\% of GPU-hours. In terms of resource profile, 41\% of simultaneously running job pairs have complementary behavior: one is compute-heavy (SM utilization $>$ 50\%) and memory-light ($<$ 40\% device memory), while the other is memory-heavy or compute-light. 
Even under a conservative 80\% combined-utilization cap, many concurrent pairs could safely share a GPU under the current exclusive-assignment policy.

\mypara{Observation 3: Short-Running Job Starvation.}
Under the existing FIFO scheduler, short jobs (62\% of all submissions) incur a median queueing delay of 47 minutes, which even exceeds their own median runtime of less than 30 minutes. This unfairness occurs because the scheduler assumes runtimes are unknown and thus cannot give precedence to short jobs over long ones. However, user submission behavior is structured enough to allow data-driven prediction: per-user runtime distributions exhibit low variability, and 78\% of users repeatedly submit jobs with similar characteristics and durations. Reliable runtime estimates make it possible to schedule short jobs first, avoid preempting nearly completed jobs, and predict the lifetime of colocated job pairs.

\subsection{Motivation}\label{subsec:bg-motivation}

The observations above suggest that underutilization arises primarily from rigid resource-management policies rather than insufficient workload demand. Improving utilization therefore requires elastic borrowing across tenants, differentiated treatment of \emph{guaranteed} and \emph{best-effort} work, runtime-aware scheduling to reduce head-of-line blocking, and interference-aware colocation to exploit complementary jobs. These mechanisms must also be tied to tenant assurance: borrowed resources should remain reclaimable, short-job acceleration should not postpone an under-served tenant, and colocation should become more conservative when it may slow quota recovery. At the same time, practical deployment demands framework-agnostic designs to avoid per-framework maintenance overhead.

\section{\SystemName Design}
\label{sec:approach}

\begin{figure}[t!]
    \centering
    \includegraphics[width=\linewidth]{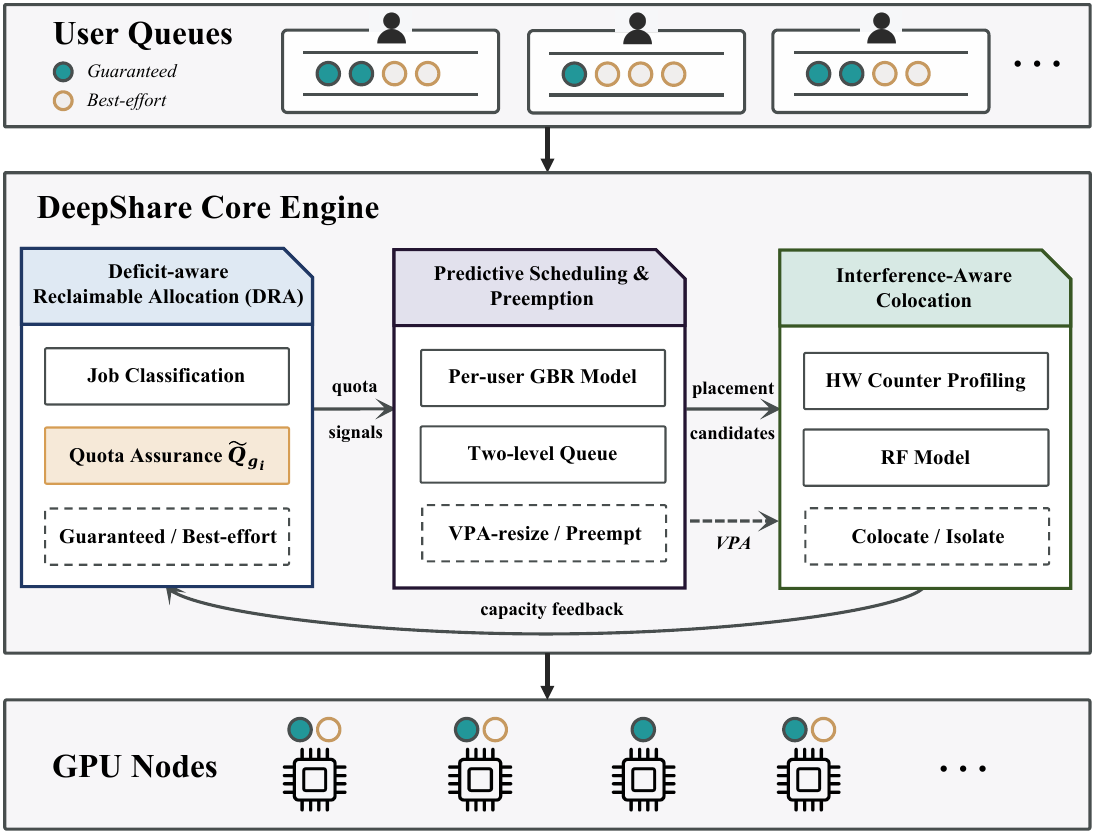}
    \caption{The architecture of \SystemName} 
    \label{fig:system_overview}
\end{figure}

\SystemName organizes scheduling decisions around a tenant-assurance state, the Quota Assurance Degree $\tilde{Q}_i(t)$, as shown in Figure~\ref{fig:system_overview}. \SystemName uses this state in three places: DRA (\S\ref{subsec:dra}) decides when idle quota can be borrowed or reclaimed, predictive scheduling and preemption (\S\ref{subsec:predictive}) order jobs and select low-cost victims, and interference-aware colocation (\S\ref{subsec:colocation}) decides when two jobs can safely share a GPU without delaying tenant recovery.

\begin{table}[t]
\centering
\caption{Notation and definitions.}
\label{tab:notation}
\scriptsize
\setlength{\tabcolsep}{3pt}
\begin{tabular}{@{}l p{0.67\columnwidth}@{}}
\toprule
\textbf{Symbol} & \textbf{Description} \\
\midrule
$K,\;G_{\mathrm{tot}}$ & Number of GPUs and total GPU capacity \\
$G_f,\;G_p$ & Free GPUs and total pending GPU demand \\
$d,\;\mathcal{J}_d$ & GPU device and its resident jobs \\
$R_j$ & GPU demand of job $j$ \\
$\mathcal{D}^P_j,\;V(d)$ & Preemption-feasible devices and victims on device $d$ \\
\midrule
$i,\;q_i$ & Tenant index and \emph{guaranteed} GPU quota \\
$A^G_i,\;D^G_i$ & Allocated and demanded \emph{guaranteed} GPUs \\
$U^G_i,\;U^B_i$ & Current \emph{guaranteed} and \emph{best-effort} GPU usage \\
$Q_i,\;\tilde{Q}_i$ & Instantaneous and EMA-smoothed QAD \\
$\eta,\;\tau_p$ & \emph{Best-effort} cap multiplier and preemption window \\
\midrule
$\mathcal{Q}^G_i,\;\mathcal{Q}^B_i$ & Tenant-level \emph{guaranteed} and \emph{best-effort} queues \\
$\mathcal{Q}^G,\;\mathcal{Q}^B$ & Cycle-local \emph{guaranteed} and \emph{best-effort} placement queues \\
$\hat{T}(j),\;\bar{T}(J)$ & Predicted remaining runtime and victim-set mean runtime \\ $C_p(j),\;\Phi(J)$ & Normalized preemption overhead and victim-set cost \\
\midrule
$\rho,\;\hat{\rho}$ & Retention ratio and predicted retention \\
$\rho_{\mathrm{tol}},\;\rho_{\min}$ & Admission tolerance and retention floor \\
$P$ & Cluster contention pressure \\
\bottomrule
\end{tabular}
\end{table}

\subsection{DRA: Deficit-aware Reclaimable Allocation}
\label{subsec:dra}

Strictly static quota enforcement causes GPUs to be underutilized when multi-tenant demand is bursty, even though DL workloads differ in how much reclaimable excess capacity they can safely tolerate, analogous to oversubscription~\cite{sun2018rose}. \SystemName introduces \textit{Deficit-aware Reclaimable Allocation} (DRA), which addresses these issues by translating quota fulfillment into a continuous control signal and separating scheduling semantics from the underlying physical placement.

\mypara{Workload classes.}
\SystemName targets DL training workloads and distinguishes between quota-backed and opportunistic execution~\cite{vavilapalli2013apache,verma2015large}. Each tenant labels submitted jobs as either \textit{guaranteed} or \textit{best-effort}. \textit{Guaranteed} jobs are charged against the quota $q_i$ and prioritized for placement backed by the quota. \textit{Best-effort} jobs are not charged against the quota and incur negligible scheduling cost; they utilize surplus capacity and remain reclaimable when \textit{guaranteed} demand is underserved.

\mypara{Quota Assurance Degree.}
For each tenant $i$, \SystemName defines the instantaneous Quota Assurance Degree (QAD) as in Eq.~\ref{eq:quota-assurance}. QAD contrasts the tenant’s current \emph{guaranteed} allocation with $\min(q_i, D^G_i(t))$, where $D^G_i(t)$ is its present \emph{guaranteed} demand. This definition prevents tenants from boosting their priority by declaring demand far above their quota, while also ensuring they are not penalized for leaving part of their quota unused. In this way, QAD measures how effectively the scheduler fulfills the \emph{guaranteed} demand that the tenant is entitled to receive.

\begin{equation}
\label{eq:quota-assurance}
Q_i(t)=
\begin{cases}
1, & D^G_i(t)=0, \\[3pt]
\displaystyle{A^G_i(t)}/{\min(q_i,D^G_i(t))}, & D^G_i(t)>0.
\end{cases}
\end{equation}

\mypara{Temporal smoothing and control.}
Binary quota mechanisms (e.g., Kubernetes ElasticQuota and Volcano~\cite{k8sschedulerplugins,volcano2024}) provide only a coarse indication of whether a tenant is within its quota. This granularity is insufficient for prioritizing recovery among under-served tenants or regulating the aggressiveness of sharing. In contrast, QAD provides a continuous signal: lower values indicate greater deficit, while smoothing mitigates transient fluctuations due to short job completions or bursty arrivals.
To stabilize this signal, \SystemName applies exponential moving average (EMA) smoothing (Eq.~\ref{eq:ema-qad}) with default $\lambda=0.3$. Under a 50 ms scheduling cycle, a sustained deficit contributes 90\% of its steady-state effect within ${\sim}350$ ms, enabling timely reclamation while filtering sub-second noise.

The smoothed QAD, $\tilde{Q}_i(t)$, drives multiple control decisions. \emph{Guaranteed} jobs are ordered by $(\tilde{Q}_i(t)\!\uparrow,\hat{T}(j)\!\uparrow)$; colocation admission is tightened when either tenant is under-served; and reclamation is prioritized before considering \emph{best-effort} preemption. \emph{Best-effort} allocations do not contribute to $A^G_i(t)$ and therefore do not inflate QAD. Preemption is triggered by placement failure.

\begin{equation}
\label{eq:ema-qad}
\tilde{Q}_i(t)=\lambda Q_i(t)+(1-\lambda)\tilde{Q}_i(t-1)
\end{equation}

\mypara{Recovery dynamics.}
When \emph{guaranteed} demand is feasible, i.e., $\sum_i \min(q_i, D^G_i(t)) \leq G_{\mathrm{tot}}$, a \emph{guaranteed} job blocked solely by \emph{best-effort} occupancy is admitted within one preemption window $\tau_p$, subject to Kubernetes Pod termination latency and a bounded \emph{best-effort} cleanup delay $\delta$. Once $A^G_i = \min(q_i, D^G_i)$, we have $Q_i = 1$, and the EMA converges geometrically at rate $(1-\lambda)$ per scheduling cycle.

Under overload, where $\sum_i \min(q_i, D^G_i(t)) > G_{\mathrm{tot}}$, full recovery is infeasible. In this regime, prioritizing tenants in ascending order of $\tilde{Q}_i$ approximates max-min fair recovery among active tenants.

\subsection{Predictive Scheduling and Preemption}
\label{subsec:predictive}
Conventional schedulers~\cite{vavilapalli2013apache,verma2015large} generally assume unknown training durations and use FIFO or static priorities, which causes head-of-line blocking. While integrating runtime prediction can alleviate this, prioritizing jobs solely by predicted time can undermine tenant-level guarantees.

\SystemName instead adopts a lexicographic policy: \emph{guaranteed} jobs are ordered first by QAD and then by the predicted remaining time. QAD determines inter-tenant recovery priority, while prediction refines intra-tenant ordering among jobs with similar assurance levels. This design improves responsiveness and reduces unnecessary preemption of \textit{best-effort} jobs that are near completion.

\mypara{Hierarchical queueing.}
\SystemName implements tenant-level admission queues and cycle-local cluster placement queues. Each cycle initializes $\mathcal{Q}^G,\mathcal{Q}^B=\emptyset$ and provisional usage $a_i^G=U_i^G$, $a_i^B=U_i^B$. Appending job $j$ immediately increases the corresponding $a_i$ by $R_j$, so the promoted batch cannot exceed $q_i$ for \emph{guaranteed} jobs or $\eta q_i$ for \emph{best-effort} jobs, with default $\eta=2$. \texttt{PlaceJobs} atomically updates actual usage after successful placement; unplaced jobs remain in their tenant queues when the cycle-local state is discarded.

At the cluster level, jobs are scheduled using a lexicographic priority. \emph{Guaranteed} jobs are considered before \emph{best-effort} jobs; within each class, jobs are ordered by $(\tilde{Q}_i(t)\!\uparrow,\hat{T}(j)\!\uparrow)$. Because QAD is the primary key, runtime predictions only refine job ordering after tenant-assurance priority has been determined. Prediction errors therefore affect only local ordering among similarly served tenants; they cannot override quota assurance, elevate a well-served tenant above an under-served one, or trigger preemption of \textit{guaranteed} jobs.

\begin{algorithm}[!b]
\caption{Scheduling Strategy}
\label{alg:scheduling}
\KwIn{Tenants $U$, cluster state}
\KwOut{Updated placements}

$\mathcal{Q}^G,\mathcal{Q}^B\gets\emptyset$\;
\ForEach{$\mathrm{tenant}~i\in U$}{
    $a_i^G\gets U_i^G$\;
    \ForEach{ $\mathrm{job}~j\in\mathcal{Q}^G_i$}{
        \If{$a^G_i+R_j\leq q_i$}{
            Append $j$ to $\mathcal{Q}^G$;\ $a_i^G\gets a_i^G+R_j$\;
        }
    }
}

Sort $\mathcal{Q}^G$ by $(\tilde{Q}_i(t)\!\uparrow,\hat{T}(j)\!\uparrow)$\;
\PlaceJobs{$\mathcal{Q}^G$, \textbf{true}}\;

\If{$\mathrm{no~ remaining}~j\in\mathcal{Q}^G \mathrm{~is~placeable}$}{
    \ForEach{$\mathrm{tenant}~i\in U$}{
        $a_i^B\gets U_i^B$\;
        \ForEach{ $\mathrm{job}~j\in\mathcal{Q}^B_i$}{
            \If{$a^B_i+R_j\leq\eta q_i$}{
                Append $j$ to $\mathcal{Q}^B$;\ $a_i^B\gets a_i^B+R_j$\;
            }
        }
    }
}

Sort $\mathcal{Q}^B$ by $(\tilde{Q}_i(t)\!\uparrow,\hat{T}(j)\!\uparrow)$\;
\PlaceJobs{$\mathcal{Q}^B$, \textit{false}}\;
\end{algorithm}

\mypara{Cost-based preemption.}
\SystemName{} never selects \textit{guaranteed} jobs as preemption victims. For incoming job $j$, the resident set $V(d)$ and preemption-feasible device set $\mathcal{D}^P_j$ are
\begin{equation}
\label{eq:preemption-device-set}
\begin{aligned}
V(d)&=\mathcal{J}_d,\\[-2pt]
\mathcal{D}^P_j&=\{d\mid \operatorname{Feasible}(j,d),\ V(d)\neq\emptyset,\\[-2pt]
&\hspace{12mm}\operatorname{class}(v)=\mathrm{BE},\ \forall v\in V(d)\}.
\end{aligned}
\end{equation}
Here, $\operatorname{Feasible}(j,d)$ means that $d$ satisfies $j$'s GPU-memory and node-level CPU/memory constraints after eviction. Therefore, reclaiming a shared GPU always preempts its complete resident set.

\SystemName estimates each job’s remaining runtime $\hat{T}(j)$ using an offline-trained per-tenant gradient boosting model\cite{friedman2001greedy}, with a cluster-wide fallback for cold-start tenants. We define the preemption cost of a victim set $J$ as Eq.~\ref{eq:preemption-cost},  where $\bar{T}(J)=|J|^{-1}\sum_{j\in J}\hat{T}(j)$ denotes the mean remaining runtime, and $C_p(j)$ captures the normalized cumulative preemption overhead of job $j$, discouraging repeated interruption of the same \textit{best-effort} job. The first term prioritizes preempting long-running jobs with limited prior disruption, while the second term penalizes fragmented victim sets, particularly those consisting of near-completion jobs (i.e., small $\bar{T}(J)$). We set $\alpha=0.5$ and $\beta=0.3$ by default.

\begin{equation}
\label{eq:preemption-cost}
\Phi(J)=\sum_{j\in J}\frac{1+\alpha C_p(j)}{\hat{T}(j)}
+\frac{\beta(|J|-1)}{\bar{T}(J)},
\end{equation}

The \texttt{SelectVictims} procedure selects
\begin{equation}
\label{eq:victim-selection}
d^*=\arg\min_{d\in\mathcal{D}^P_j}\Phi(V(d)),
\qquad J^*=V(d^*).
\end{equation}
Thus, both $\alpha$ and $\beta$ directly affect online selection. The scheduler reserves $d^*$ before eviction; if $\mathcal{D}^P_j=\emptyset$, it returns $\bot$ and leaves $j$ queued. Ranking $n$ candidate devices costs $O(n\log n)$.

In summary, Algorithm~\ref{alg:placement} considers placement options in order of increasing disruption: exclusive allocation, interference-aware colocation, CPU/memory reclamation, and finally GPU preemption.

\begin{algorithm}[t]
\caption{Interference-Aware Placement}
\label{alg:placement}
\KwIn{Queue $\mathcal{Q}_{\mathrm{in}}$, flag $\mathit{allowPreempt}$, cluster state}
\KwOut{Updated job-to-GPU assignments}

\ForEach{ $\mathrm{job}~j\in\mathcal{Q}_{\mathrm{in}} \mathrm{~in~priority~order}$}{
    $d\gets\mathrm{FindExclusiveDevice}(j)$\;
    \If{$d\neq\emptyset$}{
        Atomically assign $j$ to $d$\\ \textbf{continue}\;
    }

    $\mathit{placed}\gets\textbf{false}$\;
    $P\gets (G_p/(G_f+\epsilon))^\gamma$\;
    $\mathcal{C}\gets\{(d,j_e)\mid d$ is feasible for $j$ and hosts exactly one job $j_e\}$\;
    Sort $\mathcal{C}$ by $(\min(\hat{\rho}(j),\hat{\rho}(j_e))\!\downarrow,\hat{T}(j_e)\!\uparrow)$\;
    \ForEach{$(d,j_e)\in\mathcal{C}$}{
        $i\gets\mathrm{owner}(j)$;\ $i_e\gets\mathrm{owner}(j_e)$\;
        $\rho_{\mathrm{tol}}\gets\min(1,[\rho_{\min}+P(1-\rho_{\min})]\max(k-\tilde{Q}_i(t),k-\tilde{Q}_{i_e}(t)))$\;
        \If{$(j$ or $j_e$ is BE$)$ and $\hat{\rho}(j)\geq\rho_{\mathrm{tol}}$ and $\hat{\rho}(j_e)\geq\rho_{\mathrm{tol}}$}{
            Colocate $j$ on $d$\;
            $\mathit{placed}\gets\textbf{true}$;\ \textbf{break}\;
        }
    }
    \If{$\mathit{placed}$}{
        \textbf{continue}\;
    }

    \If{$\mathit{allowPreempt}$}{
        Shrink BE Pods via in-place resize
        \\Update per-node CPU/mem headroom\;
        $\mathit{placed}\gets\operatorname{RetryPlacement}(j)$\;
        \If{$\mathit{placed}$}{
            \textbf{continue}\;
        }
        $(J^*,d^*) \gets$ \SelectVictims{$j$}\;
        \If{$J^*\neq\bot$}{
            Preempt $J^*$ and bind $j$ to reserved $d^*$ after release\;
        }
    }
}
\end{algorithm}

\subsection{Interference-Aware Job Colocation}
\label{subsec:colocation}

Sharing GPUs can increase utilization, but a low-interference pairing is not always a good choice. If one tenant is currently running below its \emph{guaranteed} allocation, even a modest slowdown can postpone its recovery. \SystemName therefore treats colocation as an admission-control decision: a pair is scheduled together only when the predicted throughput retention is high enough given both the current cluster load and the assurance states of the tenants involved.

\mypara{Performance retention prediction.}
\SystemName estimates pairwise interference via the throughput retention ratio
$\rho = t_{\mathrm{shared}} / t_{\mathrm{excl}}$, where $t_{\mathrm{shared}}$ and $t_{\mathrm{excl}}$ denote the training throughput in colocated and exclusive execution, respectively.

\begin{figure*}[t!]
    \centering
    \includegraphics[width=\textwidth]{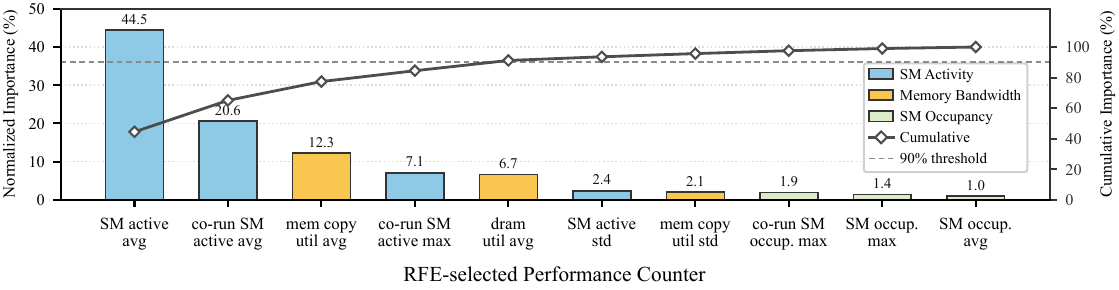}
    \caption{Feature importance scores via Recursive Feature Elimination with the Random Forest Regressor.}
    \label{fig:feature_importance}
\end{figure*}

The predictor leverages DCGM hardware telemetry, including SM activity, memory bandwidth, L2 cache behavior, PCIe traffic, tensor core utilization, DRAM throughput, and power draw, to capture both compute and memory contention effects. We use a Random Forest trained offline on isolated and colocated job profiles, balancing prediction accuracy with sub-millisecond inference latency
(\S\ref{sec:experiments}), which allows online deployment on the critical path of the scheduler.
Feature selection via Recursive Feature Elimination indicates that SM activity and memory bandwidth are the dominant predictors of interference, consistent with previous observations on GPU contention (Fig.~\ref{fig:feature_importance}).

\mypara{Dynamic tolerance.}
Let $G_p$ be total pending GPU demand and $G_f$ the number of free GPUs. Cluster contention pressure is defined as
\begin{equation}
\label{eq:contention}
P=\left(\frac{G_p}{G_f+\epsilon}\right)^{\!\gamma},
\end{equation}
where $\epsilon$ avoids division by zero. We set $\gamma=0.5$ so pressure grows sub-linearly, preventing short demand bursts from disabling all colocation. For incoming job $j$ from tenant $i$ and resident job $j_e$ from tenant $i_e$, the admission tolerance is defined in Eq.~\ref{eq:tolerance}, where $k$ denotes the maximum number of jobs colocated on a single GPU (default $k=2$).
\begin{equation}
\label{eq:tolerance}
\begin{aligned}
\rho_{\mathrm{tol}}=\min\!\bigl(&1,\,
[\rho_{\min}+P(1-\rho_{\min})] \\
&{}\times\max(k-\tilde{Q}_i(t),k-\tilde{Q}_{i_e}(t))\bigr).
\end{aligned}
\end{equation}
A pair is admitted only if it fits GPU memory and node-level CPU/memory headroom, at least one partner is \textit{best-effort}, and both $\hat{\rho}(j)$ and $\hat{\rho}(j_e)$ exceed $\rho_{\mathrm{tol}}$.
The $[\rho_{\min}+P(1-\rho_{\min})]$ term raises the tolerance when pending demand is high.
The clamp at one is deliberate: when contention is high or one of the tenants is significantly under-served, the system stops creating new colocations and the scheduler focuses on restoring exclusive capacity first.

Thus, admission behaves conservatively when the cluster is under heavy load or when a tenant’s recovery is urgent, and only becomes more permissive when the overall pressure is low and both tenants are considered sufficiently protected.

\mypara{Runtime validation and overhead.}
The predictor filters pairs before GPU sharing is enabled, but interference can change across training phases. DCGM counters feed an online retention estimator. If either job's observed retention stays below $\rho_{\mathrm{tol}}$ for $w$ consecutive samples, the pair is marked as degraded, the colocation
decision is revoked, and the \textit{best-effort} partner is preempted in the next scheduling cycle; the \textit{guaranteed} job continues running. The fixed-window rule provides bounded detection delay with one counter per pair, cheaper and more predictable than sequential detectors such as CUSUM.

Let $m$ be the number of GPUs hosting exactly one resident job. Candidate generation and prediction are $O(m)$ after pruning, and sorting costs $O(m\log m)$. 
In practice, the scheduler's node-scoring phase (\S\ref{subsec:impl-scheduler}) caps the evaluated candidates per job, keeping colocation admission within the scheduling latency budget reported in \S\ref{subsec:impl-latency}.

\section{System Implementation}
\label{sec:implementation}

The prototype contains roughly 11.3K lines of code: Go for the scheduler plugin, quota controller, and node-local DaemonSets; Python for offline-trained Random Forest interference models and
per-tenant runtime estimators; and Helm/RBAC manifests plus lightweight quota CRDs for deployment.
All mechanisms run on the cluster-management layer and do not require changes to user code or DL frameworks.

\subsection{Scheduler Critical Path}
\label{subsec:impl-scheduler}

\SystemName is packaged as an out-of-tree scheduler plugin that registers with the standard Kubernetes scheduling framework~\cite{k8sschedulerplugins}. Tenant quotas and \emph{best-effort} cap multipliers are configured through a lightweight \texttt{TenantQuota} CRD, while individual jobs specify their service class using the optional annotation \texttt{deepshare.io/class: guaranteed|best-effort}.

A lightweight quota controller reconciles \texttt{TenantQuota} objects and job-class annotations into per-tenant quota metadata, but it does not make placement decisions. The scheduler plugin keeps the real-time control loop: it maintains $Q_i(t)$ and $\tilde{Q}_i(t)$ in memory, orders queues, admits colocation pairs, and invokes preemption. The QAD state is derivable from running Pods in the informer cache; after failover, the newly elected leader reconstructs it and warm-starts the EMA with the first cycle's instantaneous QAD, avoiding extra writes or storage.

The plugin maps \SystemName's policy to five extension points. (i) \textit{Filter} keeps nodes that can host the incoming Pod on a free GPU or an eligible single-resident GPU under CPU, memory, GPU-memory, and \emph{best-effort} cap constraints. (ii) \textit{Score} ranks these candidates using the Random Forest interference model and the bilateral tolerance in \S\ref{subsec:colocation}. (iii) \textit{Reserve} atomically records device claims and cycle-local tenant reservations in the next-cycle cluster view to avoid double booking and quota over-admission. (iv) \textit{PostFilter} evaluates each feasible device's complete \emph{best-effort} victim set using Eq.~\ref{eq:preemption-cost} and reserves the selected device before eviction. (v) \textit{Permit} holds an incoming Pod only when the VPA-based CPU/memory reclamation step in Algorithm~\ref{alg:placement} has issued an in-place resize and waits until \texttt{Pod.Status.Resources} reflects the reduced allocation.

\subsection{GPU Sharing and Runtime Protection}

Spatial sharing is provided by NVIDIA MPS. One MPS control daemon is deployed per GPU in a node-local DaemonSet; it brokers client connections and sets per-client memory limits through MPS controls where supported. Because MPS multiplexes clients rather than isolating SM or memory-bandwidth contention, \SystemName treats sharing as an admission-control problem governed by the interference model in \S\ref{subsec:colocation}.

A DCGM poller samples each colocated pair. If observed retention stays below the bilateral $\rho_{\mathrm{tol}}$ for three consecutive windows, the poller raises a degradation event; the scheduler consumes it in the next cycle and preempts the \emph{best-effort} partner. A DaemonSet recovery hook clears stale MPS client contexts after abnormal Pod exits.

\mypara{In-place resize.}
Because \texttt{nvidia.com/gpu} is defined as a Kubernetes Extended Resource, its allocation cannot be modified once the Pod has been admitted.
The scheduler relies on the Pod \texttt{resize} subresource, which is enabled via \texttt{InPlacePodVerticalScaling} on our control plane, and it maintains CPU and memory headroom above the VPA recommendations of $\max(10\%,0.5\,\text{core})$ and $\max(10\%,256\,\text{MB})$, respectively.

\subsection{Scheduling Latency and Fault Tolerance}
\label{subsec:impl-latency}

End-to-end scheduling latency stays below 50\,ms per job, dominated by the Kubernetes bind round trip. Feature extraction, queue bookkeeping, and capped interference scoring together account for less than 25\,ms, keeping the latency envelope on par with Lucid~\cite{hu2023lucid}.

The plugin runs as a replicated \texttt{Deployment} with leader election through \texttt{coordination.k8s.io/Lease}. With Kubernetes default lease settings, a newly elected replica begins scheduling within one renewal period. Because preemption is expressed as Pod deletion and reconciled idempotently by the API server, no bespoke compensation logic is needed if leadership changes mid-cycle.

\section{Experiments}
\label{sec:experiments}
This section assesses \SystemName in terms of prediction accuracy, colocation efficiency, multi-tenant quota management, and deployment in a real cluster.

\begin{figure}[t!]
    \centering
    \includegraphics[width=\columnwidth]{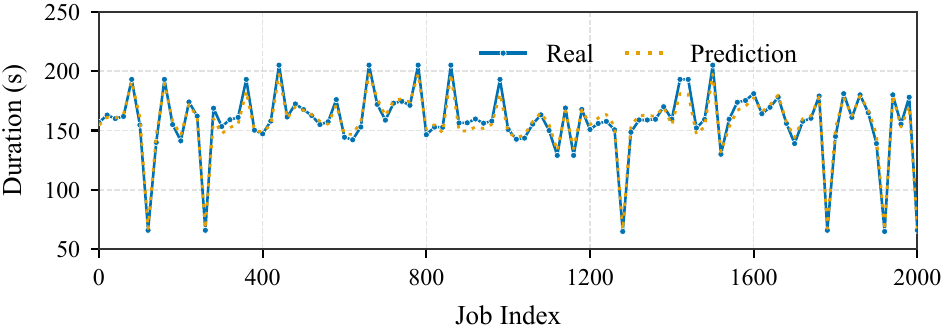}
    \caption{Predicted vs.\ actual job execution times for 2{,}000 randomly sampled jobs from the Venus dataset.}
    \label{fig:prediction_scatter}
\end{figure}

\subsection{Experiment Setup}\label{subsec:setup}

\mypara{Evaluation environments.}

We use production traces from a Kubernetes-managed cluster with 58 nodes and 219 GPUs for trace-driven evaluation. We also deploy \SystemName{} end to end on a 16-A100 Kubernetes testbed and run 50 jobs. The testbed uses Ubuntu 22.04, Kubernetes v1.35 with \texttt{InPlacePodVerticalScaling}, CUDA 12.8, and NVIDIA DCGM 3.3.8. The
58-node production cluster provides telemetry and traces for
large-scale replay, while the 16-A100 testbed hosts the complete
DeepShare prototype for end-to-end validation. This separation
distinguishes large-scale policy evaluation from implementation
feasibility under real Kubernetes scheduling and GPU sharing.

\mypara{Profiling workloads.}
Table~\ref{tab:profiling_workloads} summarizes the representative
single-GPU training workloads used for
interference profiling. The suite covers diverse model structures
and resource behaviors, while synthetic inputs remove data-loading
and storage I/O variability.

\begin{table}[t]
\centering
\caption{Representative single-GPU training workloads used for interference profiling.}
\label{tab:profiling_workloads}
\scriptsize
\setlength{\tabcolsep}{2.4pt}
\renewcommand{\arraystretch}{1.02}
\resizebox{\columnwidth}{!}{%
\begin{tabular}{@{}lllc@{}}
\toprule
\textbf{Domain} &
\textbf{Workload} &
\textbf{Input configuration} &
\textbf{Batch} \\
\midrule

CV
& ResNet-18~\cite{he2016deep}
& $224{\times}224$ classification
& 128 \\

CV
& ResNet-20~\cite{he2016deep}
& $32{\times}32$ classification
& 512 \\

CV
& MobileNetV3-Small~\cite{howard2019searching}
& Lightweight image classification
& 256 \\

CV
& ResNet-50~\cite{he2016deep}
& $224{\times}224$ classification
& 48 \\

3D
& PointNet-style~\cite{qi2017pointnet}
& 1,024 points per sample
& 96 \\

GAN
& DCGAN-style~\cite{radford2016unsupervised}
& Adversarial image generation
& 64 \\

RL
& Actor--Critic-style~\cite{mnih2016asynchronous}
& Policy and value optimization
& 4,096 \\

NLP
& BERT-style MLM~\cite{devlin2019bert}
& MLM, sequence length 128
& 48 \\

Speech
& DeepSpeech2-style~\cite{amodei2016deep}
& Conv--BiGRU with CTC
& 16 \\

RecSys
& NCF-style~\cite{he2017neural}
& Embedding with MLP
& 8,192 \\

NLP
& Transformer seq2seq~\cite{vaswani2017attention}
& Encoder--decoder, sequence length 64
& 48 \\

CV
& U-Net-style~\cite{ronneberger2015unet}
& $128{\times}128$ image translation
& 8 \\

NLP
& LSTM language model~\cite{hochreiter1997long}
& 2-layer LM, sequence length 192
& 128 \\

\bottomrule
\end{tabular}%
}
\end{table}

\mypara{Workloads.}
We used two real-world datasets, summarized in Table~\ref{tab:datasets}.
The Venus trace~\cite{hu2021characterization} contains large-scale GPU datacenter scheduling and utilization records from September 2020, providing diverse public workloads for reproducible evaluation. The internal trace preserves user quotas, submissions, and allocation policies from our institutional cluster; its longer average runtime (36{,}887\,s vs.\ 5{,}419\,s) stresses scheduling under long-lived occupancy.

\begin{table}[t!]
\centering
\caption{Characteristics of the simulation datasets.}
\label{tab:datasets}
\begin{tabular}{llrc}
\toprule
\textbf{Dataset} & \textbf{Source} & \textbf{Job Count} & \textbf{Avg. Execution Time} \\
\midrule
Venus~\cite{hu2021characterization} & Public dataset & 23{,}859 & 5{,}419\,s \\
Internal & Internal cluster & 3{,}200 & 36{,}887\,s \\
\bottomrule
\end{tabular}
\end{table}

\mypara{Trace replay methodology.}
We implemented a trace-driven simulator that reproduces \SystemName's two-level queues, DRA module, preemption, and interference-aware colocation. 
It supports FIFO, SJF, QSSF, Tiresias, Lucid, and \SystemName's strategies. 

For the Venus dataset, which does not include per-tenant quota assignments, we synthesize a multi-tenant quota configuration as follows: jobs are partitioned into 12 virtual clusters by user ID hash, and each virtual cluster receives a fixed quota proportional to its historical GPU-hour share (range: 4--32 GPUs), mirroring the quota distribution observed in our university cluster (\S\ref{subsec:bg-empirical}).

\begin{figure*}[t!]
    \centering
    \includegraphics[width=\textwidth]{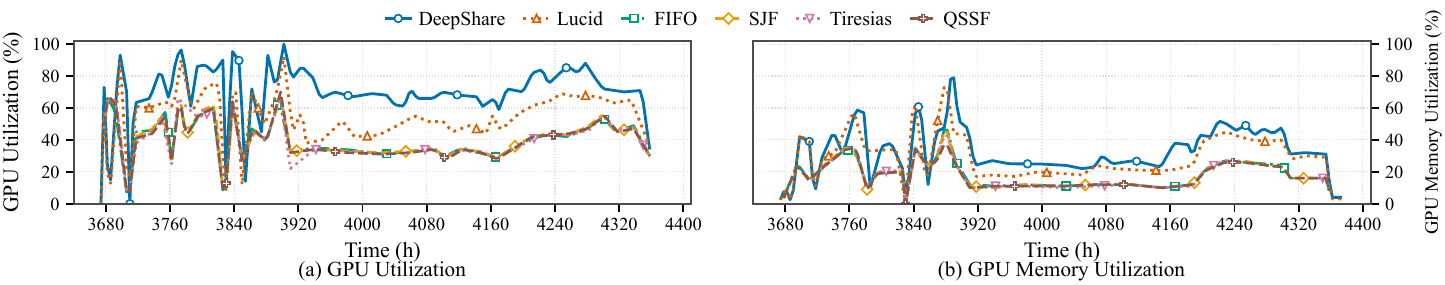}
    \caption{GPU and GPU memory utilization over time.}
    \label{fig:colocation_heatmap}
\end{figure*}

\mypara{Metrics.}
We evaluated the performance of our scheduling strategies using the following metrics:
(1)~\emph{Job Completion Time} (JCT): time from job submission to completion, encompassing both queueing and execution time; (2)~\emph{Average Queueing Delay}: average time a job spends in the queue before being scheduled; (3)~\emph{Makespan}: total time from the first submission to the last completion; (4)~\emph{Cluster GPU Utilization}: average percentage of GPU compute resources used; (5)~\emph{GPU Memory Utilization}: average percentage of GPU memory used; and (6)~\emph{Quota Assurance Degree}: as defined in Eq.~\ref{eq:quota-assurance}, measuring SLA fulfillment.

\mypara{Baselines.}
We compare \SystemName against five baselines: {FIFO}, {SJF}, {QSSF}~\cite{hu2021characterization}, {Tiresias}~\cite{gu2019tiresias}, and {Lucid}~\cite{hu2023lucid}. FIFO, SJF, QSSF, and Tiresias represent scheduling-only policies with different ordering assumptions, while Lucid is the strongest non-intrusive sharing baseline. All baselines use identical quota assignments and workload traces. 

\subsection{Job Execution Time Prediction Accuracy}\label{subsec:exp-prediction}

We assessed the accuracy of our job execution time prediction algorithm in
comparison to Lucid~\cite{hu2023lucid} using the Venus dataset.
Table~\ref{tab:prediction-accuracy} summarizes the key metrics.

\begin{table}[t!]
\centering
\caption{Accuracy metrics of job execution time prediction algorithms on the Venus dataset.}
\label{tab:prediction-accuracy}
\begin{tabular}{lcc}
\toprule
\textbf{Algorithm} & \textbf{MAPE (\%)} & \textbf{$R^2$ Score} \\
\midrule
Lucid & 68.72 & 0.6413 \\
\SystemName{} & \textbf{31.84} & \textbf{0.7286} \\
\bottomrule
\end{tabular}
\end{table}

\SystemName{} reduces MAPE from 68.72\% to 31.84\% (a 2.16$\times$ error reduction) and raises $R^2$ from 0.6413 to 0.7286 on the Venus dataset (Table~\ref{tab:prediction-accuracy}). Fig.~\ref{fig:prediction_scatter} confirms tight predicted-vs.-actual alignment, with accuracy highest for users with $\geq$50 historical submissions (MAPE\,$<$\,25\%) and graceful degradation for cold-start users via the cluster-wide fallback (MAPE\,$<$\,60\%).

\subsection{Colocation Strategy Performance}\label{subsec:exp-colocation}

We evaluated \SystemName{}'s interference-aware colocation strategy against non-sharing (FIFO, SJF, QSSF, Tiresias) and sharing (Lucid) baselines.

\mypara{Resource utilization.}
Table~\ref{tab:gpu-utilization} compares GPU and memory utilization across all strategies.

\begin{table}[t!]
\centering
\caption{Comparative GPU resource utilization across scheduling strategies.}
\label{tab:gpu-utilization}
\footnotesize
\resizebox{\columnwidth}{!}{%
\begin{tabular}{lcccccc}
\toprule
\textbf{Metric} & \textbf{FIFO} & \textbf{SJF} & \textbf{QSSF} & \textbf{Tiresias} & \textbf{Lucid} & \textbf{\SystemName{}} \\
\midrule
GPU Util (\%) & 39.64 & 40.00 & 39.40 & 39.27 & 54.52 & \textbf{70.58} \\
Mem Util (\%) & 17.94 & 17.72 & 17.42 & 17.48 & 28.74 & \textbf{32.67} \\
\bottomrule
\end{tabular}%
}
\end{table}

\SystemName{} achieves 70.58\% average GPU utilization and 32.67\% memory utilization, representing a 29.5\% relative improvement in GPU utilization over Lucid and a 78.1\% improvement over FIFO. The utilization gain over Lucid is non-trivial because both systems perform colocation; the difference stems from three factors: (i)~\SystemName{}'s more accurate interference model ($R^2 = 0.902$ vs.\ Lucid's simpler scoring approach) enables it to accept more colocation pairs that Lucid conservatively rejects; (ii)~the dynamic tolerance mechanism aggressively colocates jobs when contention is low (i.e., when $P=\left(\frac{G_p}{G_f+\epsilon}\right)^{\!\gamma}\approx 0$), exploiting off-peak periods that static thresholds cannot adapt to; and (iii)~the DRA module channels additional \emph{best-effort} jobs into idle capacity, increasing the pool of colocation candidates.

\begin{figure}[t!]
    \centering
    \includegraphics[width=\columnwidth]{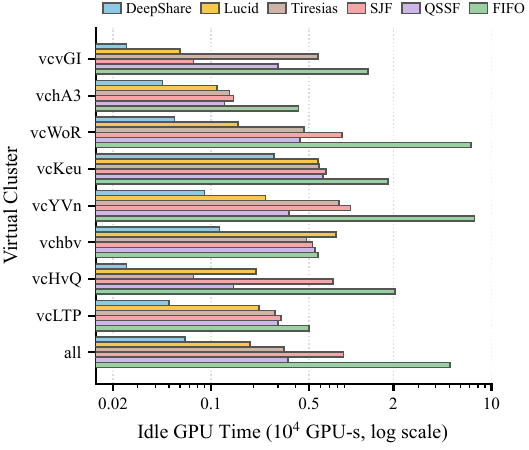}
    \caption{Idle GPU time across different Venus virtual-cluster
configurations.}
    \label{fig:idle_gpu_time}
\end{figure}

\begin{figure}[t!]
    \centering
    \includegraphics[width=\columnwidth]{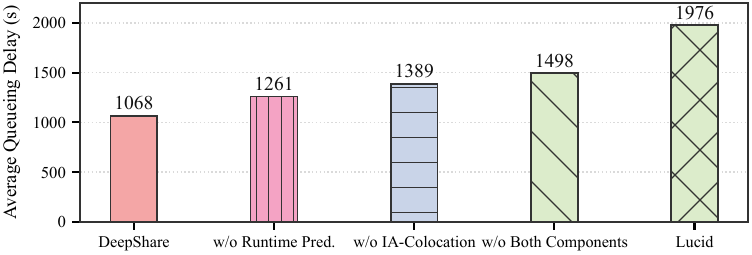}
    \caption{Ablation of average queueing delay.}
    \label{fig:ablation}
\end{figure}

The improvement in memory utilization over Lucid is more modest (13.7\%) because GPU memory is a hard constraint that limits the scope of sharing.
Fig.~\ref{fig:colocation_heatmap} shows the temporal dynamics of GPU utilization across strategies. \SystemName{} maintains a consistently higher and more stable utilization over time, with fewer periods of underutilization.

\mypara{Job completion time and queueing delay.}
Table~\ref{tab:performance-comparison} presents the JCT and queueing delay results. \SystemName{} reduces average queueing delay by 46\% over Lucid (1,068 s vs. 1,976 s) and by 98\% over FIFO. This is especially important for short exploratory and opportunistic jobs, whose responsiveness is often dominated by waiting time. The JCT improvement over Lucid is smaller (6.3\%) because execution time is mainly workload-dependent and remains broadly comparable across schedulers for the same job set.

Accordingly, \SystemName{} should be interpreted primarily as improving admission responsiveness and tenant-level resource
assurance rather than accelerating the computation of an already-running job. The end-to-end benefit is therefore most visible for exploratory and short-running jobs whose latency is dominated by queueing. For long-running training jobs, execution time dominates JCT, so the relative JCT reduction is naturally smaller even when the absolute waiting-time reduction remains substantial.

\begin{figure}[t!]
    \centering
    \includegraphics[width=\columnwidth]{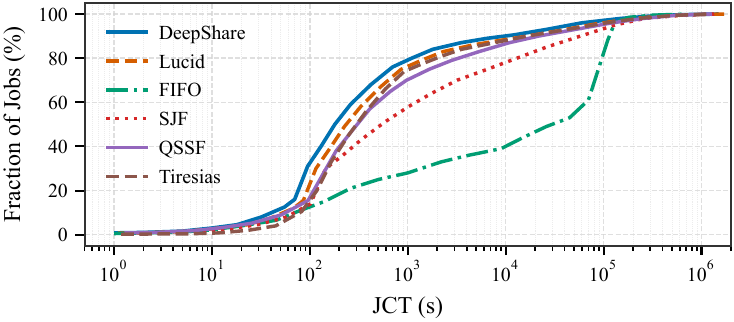}
    \caption{CDF of job completion times.}
    \label{fig:colocation_cdf}
\end{figure}

\begin{table}[t!]
\centering
\caption{Performance comparison of scheduling policies (all values in seconds).}
\label{tab:performance-comparison}
\footnotesize
\resizebox{\columnwidth}{!}{%
\begin{tabular}{lcccccc}
\toprule
\textbf{Metric} & \textbf{FIFO} & \textbf{SJF} & \textbf{QSSF} & \textbf{Tiresias} & \textbf{Lucid} & \textbf{\SystemName{}} \\
\midrule
JCT & 64{,}314 & 21{,}094 & 15{,}550 & 14{,}684 & 13{,}877 & \textbf{13{,}001} \\
Queueing Delay & 52{,}546 & 9{,}325 & 3{,}090 & 2{,}900 & 1{,}976 & \textbf{1{,}068} \\
\bottomrule
\end{tabular}%
}
\end{table}

The CDF of job completion times (Fig.~\ref{fig:colocation_cdf}) shows that \SystemName{} achieves faster completion across all percentiles. Notably, the improvement is most pronounced at the tail (P95 and P99), where \SystemName{} reduces tail JCT by 23\% compared to Lucid, indicating that it effectively prevents worst-case scenarios for long-queued jobs.

\SystemName consistently achieves the lowest idle GPU time across all tested Venus cluster configurations, shown in Fig.~\ref{fig:idle_gpu_time}. Averaged over the eight configurations, it reduces idle GPU time by 71.0\% relative to Lucid and 96.8\% relative to FIFO; in the aggregate “all” configuration, idle GPU time further drops to $0.065\times 10^4$ GPU-s, compared with $0.190\times 10^4$ for Lucid and $5.05\times 10^4$ for FIFO.

\mypara{Ablation study.}
Fig.~\ref{fig:ablation} isolates the contributions of runtime prediction and
interference-aware colocation to average queueing delay. The full \SystemName{} strategy reduces average queueing delay by 45.9\% relative to Lucid ($p < 0.01$, Wilcoxon signed-rank test). Removing runtime prediction increases queueing delay by 18.4\%, while removing interference awareness increases it by 30.1\%, showing that both components matter and that colocation contributes the larger share.

Their interaction with DRA is evaluated separately in Fig.~\ref{fig:dra_synergy}. In all configurations, QAD remains the primary control signal,
preventing short-job optimization and aggressive sharing from
overriding tenant recovery.

\subsection{Multi-Tenant Quota Management}\label{subsec:exp-quota}

\begin{figure}[t!]
    \centering
    \includegraphics[width=\columnwidth]{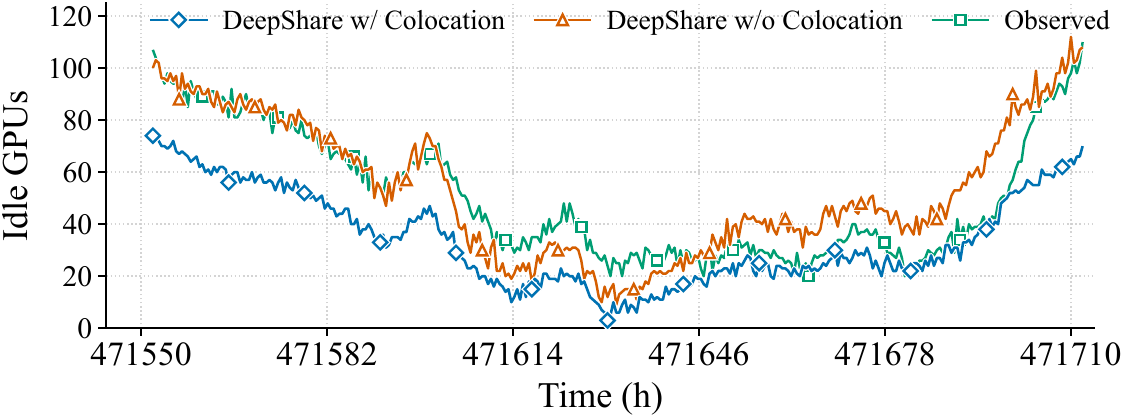}
    \caption{Idle GPU counts over time.}
    \label{fig:idle_gpus}
\end{figure}

\begin{figure}[t!]
    \centering
    \includegraphics[width=\columnwidth]{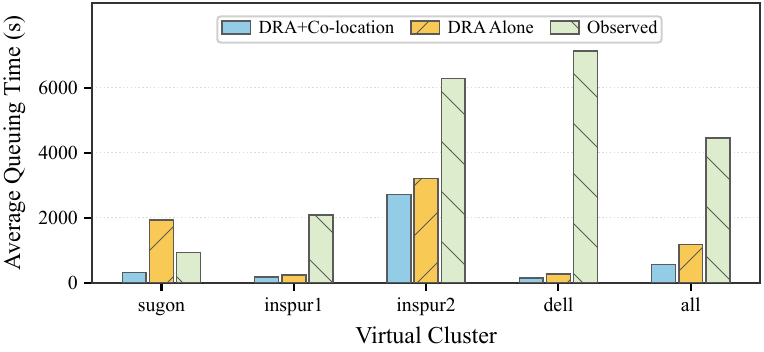}
    \caption{Queueing delay with DRA alone and with DRA plus
interference-aware colocation. The further 31\% reduction shows the
synergy between elastic quota allocation and colocation.}
    \label{fig:dra_synergy}
\end{figure}

We evaluate DRA's impact on fairness and queueing delay against fixed-quota schedulers, Tiresias, and real-world system traces (\textit{Real}) from the internal cluster. The \textit{Observed} series in Figs.~\ref{fig:idle_gpus} and~\ref{fig:dra_synergy} denotes this original internal-cluster behavior and corresponds to \textit{Real} in Table~\ref{tab:quota-management}. We report average instantaneous QAD $Q_i(t)$ to measure raw quota satisfaction, and use smoothed $\tilde{Q}_i(t)$ for cycle-level QoS compliance because it matches the scheduler's control signal.
Table~\ref{tab:quota-management} presents the key metrics.

\begin{table}[t!]
\centering
\caption{Quota guarantee and queueing delay across strategies.}
\label{tab:quota-management}
\begin{tabular}{lcc}
\toprule
\textbf{Strategy} & \textbf{Avg. $Q_i(t)$} & \textbf{Avg. Queueing Delay (s)} \\
\midrule
FIFO & 1.00 & 98{,}000 \\
SJF & 1.00 & 42{,}387 \\
QSSF & 1.00 & 97{,}271 \\
Tiresias & 0.15 & 86{,}754 \\
Real & 0.75 & 4{,}466 \\
\SystemName{} & \textbf{1.00} & \textbf{1{,}168} \\
\bottomrule
\end{tabular}
\end{table}

Among the evaluated strategies, \SystemName{} is the only one that achieves an average $Q_i(t)$ of 1.0 while maintaining low queueing delay (1{,}168\,s). Fixed-quota strategies (FIFO, SJF, QSSF) trivially achieve average $Q_i(t)$ of 1.0 but at the cost of extremely high queueing delays (36--84$\times$ higher). Tiresias improves scheduling efficiency but severely violates quota guarantees (average $Q_i(t)$ of 0.15), making it unsuitable for multi-tenant environments with SLAs. The real system traces show a compromise (average $Q_i(t)$ of 0.75, delay 4{,}466\,s) that neither fully satisfies quotas nor minimizes delay.

DRA's \emph{elastic quota mechanism} makes this possible: tenants opportunistically use idle capacity, while QAD keeps borrowed capacity reclaimable when \emph{guaranteed} demand becomes under-served. This decouples quota compliance from utilization, so the two are no longer in tension.

Fig.~\ref{fig:idle_gpus} shows the time-series of idle GPU counts. \SystemName{} maintains consistently fewer idle GPUs, confirming that its elastic allocation effectively fills resource gaps.

\mypara{QoS compliance analysis.}
A scheduling cycle is compliant for tenant~$i$ if $\tilde{Q}_i(t) \geq 0.95$. In the physical deployment (\S\ref{subsec:exp-physical}), 93\% of tenant-cycle pairs meet this threshold. The remaining 7\% of violations concentrate during two scenarios: (i)~burst arrivals where multiple tenants temporarily exceed aggregate capacity, and (ii)~transient periods immediately following preemption events, before $\tilde{Q}_i(t)$ recovers.
In the Venus simulation, the median per-tenant $\tilde{Q}_i(t)$ is 0.98 (IQR 0.94--1.00), and no tenant experiences $\tilde{Q}_i(t) < 0.85$ for more than 2\% of scheduling cycles, confirming per-tenant (not just aggregate) compliance.

\begin{figure*}[t!]
    \centering
    \includegraphics[width=\textwidth]{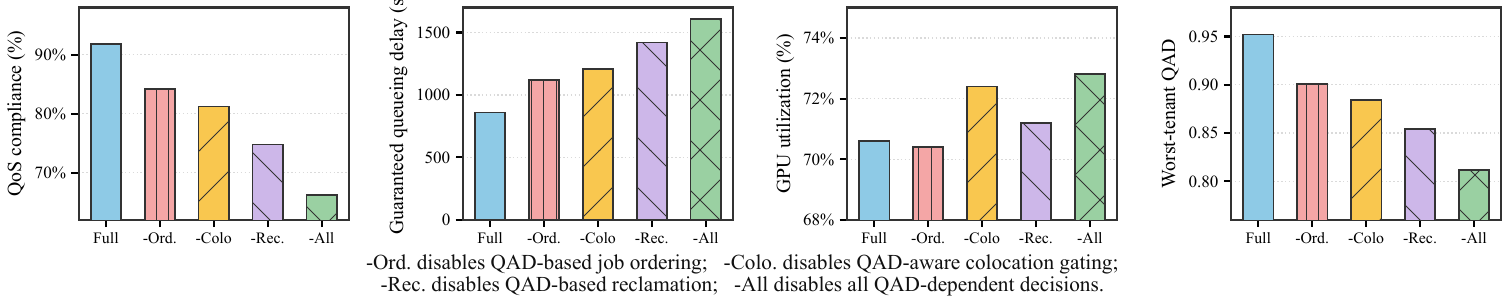}
    \caption{QAD ablation across ordering, colocation gating, and reclamation.}
    \label{fig:qad_ablation}
\end{figure*}

\mypara{QAD ablation.} To isolate the role of QAD, we ablate its use in job ordering (-Ord.), colocation gating (-Colo.), \emph{best-effort} reclamation (-Rec.), and all QAD-dependent decisions (-All), while keeping the remaining DeepShare mechanisms enabled. As shown in Fig. \ref{fig:qad_ablation}, removing QAD consistently weakens tenant assurance. In the QAD-ablation workload, Full \SystemName{} achieves about 92\% QoS compliance and a worst-tenant average QAD of about 0.95, whereas -All reduces them to about 66\% and 0.81, respectively. \emph{Guaranteed}-job queueing delay also increases from below 900 seconds in Full \SystemName{} to roughly 1,600 seconds in -All. In contrast, GPU utilization remains comparable and can even be slightly higher without QAD, because the scheduler admits borrowing and colocation more aggressively. This confirms that QAD is not primarily a utilization booster; rather, it is the control signal that prevents prediction, borrowing, and colocation from sacrificing under-served tenants.

\mypara{Overload behavior.}
When aggregate \emph{guaranteed} demand exceeds cluster capacity (approximately 8\% of Venus peak-hour cycles), \SystemName{} degrades gracefully: Algorithm~\ref{alg:scheduling} prioritizes the most under-served tenants by ascending $\tilde{Q}_i(t)$ order. The worst-case per-tenant $\tilde{Q}_i(t)$ during overload is 0.72, with recovery to $\tilde{Q}_i(t)\geq0.95$ within five cycles ($\sim$250\,ms) after the spike subsides. \emph{Best-effort} jobs absorb most of the overload cost, with a 2.1$\times$ increase in queueing delay versus only a 14\% increase for \emph{guaranteed} jobs, confirming the intended service differentiation.

\mypara{Synergy between DRA and colocation.} Fig.~\ref{fig:dra_synergy} evaluates their
combined effect and shows that adding interference-aware colocation
to DRA yields a further 31\% reduction in queueing delay relative
to DRA alone. This synergy arises because colocation increases the cluster's usable capacity by placing two jobs per GPU where possible, while QAD prevents those placements from delaying recovery of under-served tenants.

\begin{figure}[t!]
    \centering
    \includegraphics[width=\columnwidth]{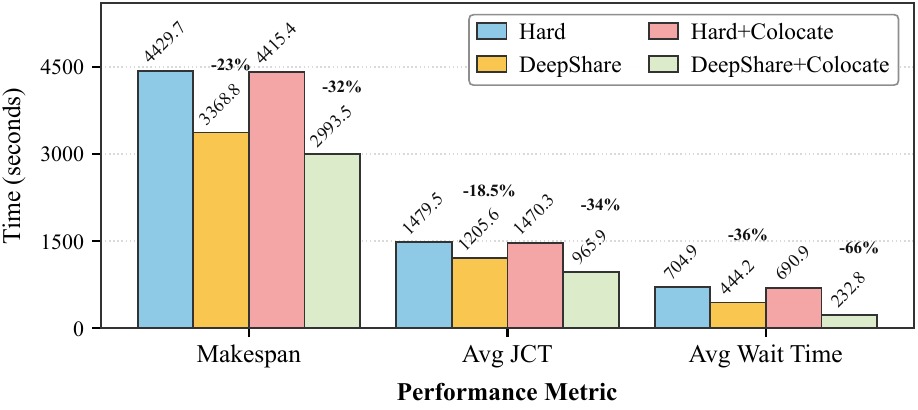}
    \caption{Physical cluster results. \SystemName{} with colocation achieves the best performance across all metrics, confirming practical deployability.}
    \label{fig:physical_bar}
\end{figure}

\subsection{Evaluation on a Physical Cluster}\label{subsec:exp-physical}

To validate practical deployability, we evaluated \SystemName{} on our Kubernetes-managed testbed. We ran 50 jobs following internal cluster patterns (1-GPU: 77\%, 2-GPU: 15\%, 4-GPU: 8\%) and measured JCT, queueing delay, and makespan.

Fig.~\ref{fig:physical_bar} presents the results under four configurations: \textbf{Hard} (fixed quota, no colocation), \textbf{Hard+Colocate} (fixed quota with colocation), \textbf{\SystemName{}} (DRA without colocation), and \textbf{\SystemName{}+Colocate} (full system).

With DRA alone, makespan decreased from 4{,}429.7\,s to 3{,}368.8\,s (24.0\% reduction), average JCT from 1{,}479.5\,s to 1{,}205.6\,s (18.5\% reduction), and queueing delay from 704.9\,s to 444.2\,s (37.0\% reduction). With the full system (DRA + Colocation), compared with Hard+Colocate, makespan was reduced by 32\%, average JCT by 34\%, and queueing delay by 66\% (from 690.9\,s to 232.8\,s). The full system processes the same batch of 50 jobs in 67.6\% of the time required by the Hard baseline, corresponding to approximately 1.48$\times$ the throughput on the physical testbed.

The physical cluster improvements are slightly lower than simulation results, which is expected due to the smaller scale (16 GPUs vs.\ simulated hundreds) limiting the opportunities for elastic resource redistribution. Nevertheless, the consistent direction and magnitude of improvements across both environments confirm the robustness of \SystemName{}'s design.

\subsection{Sensitivity Analysis}\label{subsec:sensitivity}

We analyze the robustness of \SystemName{} to its two most influential parameters.

\mypara{Dynamic tolerance baseline ($\rho_{\min}$).}
We varied $\rho_{\min}$ from 0.5 to 0.95 and measured GPU utilization and average per-job degradation of colocated jobs. At $\rho_{\min} = 0.5$ (aggressive), GPU utilization increases to 71.2\% but average degradation reaches 18\%, causing a net negative impact on JCT. At $\rho_{\min} = 0.7$ (default), the system achieves 70.58\% utilization with $<$8\% average degradation. At $\rho_{\min} = 0.9$ (conservative), utilization drops to 55.3\% as colocation opportunities become too restricted. The default thus provides a robust balance between cluster efficiency and individual job impact.

\mypara{Preemption cost weight ($\alpha$).}
We varied $\alpha$ from 0 to 2.0. Higher values reduce repeated preemptions of the same job but may cause suboptimal resource reclamation. We find $\alpha \in [0.3, 0.8]$ consistently near-optimal across both datasets, with default $\alpha = 0.5$. The multi-victim penalty $\beta$ exhibits similar robustness across $[0.1, 0.6]$.

\subsection{Scalability and Limitations}\label{subsec:discussion}

Victim-set and colocation candidate ranking cost $O(n\log n)$ and $O(m\log m)$, respectively, while interference inference costs $O(1)$ per candidate pair. Extending \SystemName{} to heterogeneous accelerators requires retraining the interference model on per-device profiling data.

Our interference predictor is trained on the diverse profiling
workloads summarized in Table~\ref{tab:profiling_workloads},
covering computer vision, 3D perception, generative modeling,
reinforcement learning, NLP, speech recognition, recommendation,
and machine translation. Rather than relying on model-specific identifiers, it uses transferable DCGM hardware counters, such as SM activity and memory bandwidth. Variations in unseen architectures, training phases, batch sizes, and GPU types are therefore reflected in the runtime signals observed by the predictor. The physical testbed (16 GPUs) is smaller than production clusters; results are most directly applicable to departmental-scale clusters (tens to low hundreds of GPUs), while simulation on 23{,}859 jobs provides evidence at larger scale.

To guard against incorrect predictions, online retention validation
continuously checks colocated pairs and revokes a colocation when
the observed retention falls below the admission threshold. This
fallback affects only the reclaimable \textit{best-effort} partner and does
not compromise \textit{guaranteed} jobs.

\section{Related Work}
\label{sec:related}

\mypara{GPU cluster scheduling and workload characterization.}
Large-scale traces have revealed persistent underutilization in multi-tenant GPU clusters~\cite{jeon2019analysis, weng2022mlaas, hu2024characterization}.
Scheduling policies address this from different angles:
Tiresias~\cite{gu2019tiresias} prioritizes jobs via a multi-level feedback queue without runtime knowledge;
Chronus~\cite{gao2021chronus} targets deadline-aware scheduling;
Pollux~\cite{qiao2021pollux} co-adapts batch sizes and allocations for goodput maximization.
Other recent policies target notebook reclamation~\cite{carver2026notebookos}, serverless ML~\cite{wu2023ceScaling}, non-linear multi-GPU scalability~\cite{butt2020marble}, and heterogeneous GPU placement~\cite{butt2025rltune}.
Maestro~\cite{wang2026maestro} further explores workload-aware cross-cluster scheduling for LLM-based multi-agent systems by jointly considering stage-level execution costs, model readiness, KV-cache feasibility, and network latency.

\mypara{Interference-aware GPU sharing and colocation.}
Gandiva~\cite{xiao2018gandiva} and AntMan~\cite{xiao2020antman} pioneer DL job colocation with framework-level and asymmetric memory-scaling mechanisms, respectively.
TGS~\cite{wu2023transparent} achieves transparent sharing via driver-level rate control, while Orion~\cite{strati2024orion} partitions SMs and memory at the kernel level for inference.
Other systems address GPU fraction management~\cite{yeh2020kubeshare} and memory-layer multiplexing~\cite{yu2019salus}.
On the modeling side, SCHEDTUNE~\cite{butt2022schedtune} profiles interference on heterogeneous GPUs,
Jacquet~et~al.~\cite{jacquet2026untangling} characterize per-job GPU power under sharing, and
ISACPP~\cite{liu2025isacpp} advances interference prediction with graph attention networks at 50--200\,ms latency per pair.

\mypara{Fairness, quota management, preemption, and runtime prediction.}
Themis~\cite{mahajan2020themis}, ASTRAEA~\cite{ye2021astraea}, and Shockwave~\cite{zheng2023shockwave} formalize various fairness objectives but do not differentiate production from \emph{best-effort} workloads or integrate colocation.
Gavel~\cite{narayanan2020heterogeneity} optimizes max-min fairness over effective throughput but assumes exclusive GPU placement.
HiveD~\cite{zhao2020hived} guarantees resources via static cell partitioning;
Sia~\cite{jayaram2023sia} optimizes goodput across heterogeneous GPUs but leaves colocation and quota out of scope.
Other work addresses priority backfilling for HPC~\cite{gainaru2025priority} and MPI process malleability~\cite{iserte2026resource}.
In the Kubernetes ecosystem, the ElasticQuota plugin~\cite{k8sschedulerplugins} and Volcano~\cite{volcano2024} provide namespace-level min/max quotas with binary reclamation.
For preemption, REEF~\cite{han2022reef} targets microsecond kernel-level preemption for inference,
GPREEMPT~\cite{fan2025gpreempt} generalizes GPU preemptive mechanisms,
GFS~\cite{duan2026gfs} forecasts organizational demand using checkpoint recency, and
Parcae~\cite{duan2024parcae} migrates LLM training proactively.
On runtime prediction, Optimus~\cite{peng2018optimus} fits loss curves online and ElasticFlow~\cite{gu2023elasticflow} scales workers by predicted throughput.

\section{Conclusion}
\label{sec:conclusion}

This paper presented \SystemName{}, an assurance-driven resource management framework for multi-tenant GPU clusters that coordinates elastic quota regulation, predictive scheduling, and interference-aware colocation through the Quota Assurance Degree $\tilde{Q}_i(t)$.
Trace-driven simulation on 23{,}859 Venus jobs and 3{,}200 internal jobs shows 70.58\% average GPU utilization (29.5\% above Lucid) and 46\% lower queueing delay than Lucid, while a 16-GPU Kubernetes deployment confirms a 34\% JCT reduction and 93\% tenant-cycle QoS compliance. Combining DRA with colocation further reduces queueing delay by 31\% beyond DRA alone.
Future work includes broader validation and model adaptation across heterogeneous GPU generations, as well as support for larger distributed training jobs.

\section*{Acknowledgment}
This work was supported in part by National Key R\&D Program of China (Grant No. 2024YFB4505604), in part by the National Natural Science Foundation of China (Grant No. 62402024), in part by Beijing Natural Science Foundation (No. L241050), in part by the Fundamental Research Funds for the Central Universities.

\IEEEtriggeratref{51}
\bibliographystyle{IEEEtran}
\bibliography{refs}

\end{document}